
\newcount\eqnumber
\eqnumber=1
\def\chaphead{}

\def\new{\hbox{(\chaphead\the\eqnumber}\global\advance\eqnumber by 1}
\def\ref#1{\advance\eqnumber by -#1 (\chaphead\the\eqnumber
     \advance\eqnumber by #1 }
\def\first{\hbox{(\chaphead\the\eqnumber{a}}\global\advance\eqnumber by 1}
\def\last{\advance\eqnumber by -1 \hbox{(\chaphead\the\eqnumber}\advance
     \eqnumber by 1}
\def\eq#1{\advance\eqnumber by -#1 equation (\chaphead\the\eqnumber
     \advance\eqnumber by #1}
\def\eqnam#1{\xdef#1{\chaphead\the\eqnumber}}


\def\eqt#1{Eq.~({{#1}})}

\def\halfspace{\baselineskip=1.5\normalbaselineskip}
\def\doublespace{\baselineskip=2\normalbaselineskip}
\def\tcar{\futurelet\next\testnextcar}
\def\testnextcar{\ifhmode\ifcat\next.\else\ \fi\fi}

\def\Dt{\spose{\raise 1.5ex\hbox{\hskip4pt$\mathchar"201$}}} 
\def\Rf{\normalbaselines\parindent=0pt\medskip\hangindent=3pc\hangafter=1}
\def\a37{\hbox{$^{37}$Ar}}
\def\ar37{\hbox{$^{37}$Ar}}
\def\argon40{\hbox{$^{40}$Ar}}
\def\as71{\hbox{$^{71}$As}}
\def\b8{\hbox{$^{8}$B}}
\def\be7{\hbox{$^{7}$Be}}
\def\beril8{\hbox{$^8$Be}}
\def\boron11{\hbox{$^{11}$B}}

\def\br81{\hbox{$^{81}$Br}}
\def\carb12{\hbox{$^{12}$C}}
\def\ca37{\hbox{$^{37}$Ca}}
\def\car11{\hbox{$^{11}$C}}
\def\carbon13{\hbox{$^{13}C$}}

\def\chr51{\hbox{$^{51}$Cr}}
\def\cl37{\hbox{$^{37}$Cl}}
\def\cm2{\hbox{cm$^{2}$}}

\def\cu65{\hbox{$^{65}$Cu}}
\def\day-1{\hbox{day$^{-1}$}\tcar}

\def\dt{\spose{\raise 1.0ex\hbox{\hskip2pt$\mathchar"201$}}} 
\def\f17{\hbox{$^{17}$F}}

\def\ga71{\hbox{$^{71}$Ga}}
\def\gacl3{\hbox{GaCl$_3$}}
\def\ge71{\hbox{$^{71}$Ge}}
\def\gecl4{\hbox{GeCl$_4$}}
\def\geh4{\hbox{GeH$_4$}}

\def\h2{\hbox{$^{2}$H}}
\def\he3{\hbox{$^{3}$He}}

\def\helium4{\hbox{$^{4}$He}}

\def\hy1{\hbox{$^{1}$H}}
\def\ind115{\hbox{$^{115}$In}}
\def\iodine127{\hbox{$^{127}$I}}
\def\k37{\hbox{$~^{37}$K}}
\def\k40{\hbox{$^{40}$K}}
\def\kay40{\hbox{$~^{40}$K}}

\def\kr81{\hbox{$^{81}$Kr}}

\def\li7{\hbox{$^{7}$Li}}
\def\lithium8{\hbox{$^{8}$Li}}

\def\mo98{\hbox{$^{98}$Mo}}

\def\n13{\hbox{$^{13}$N}}
\def\nitrogen14{\hbox{$^{14}$N}}

\def\o15{\hbox{$^{15}$O}}
\def\ok2{\hbox{$~^{2}$K}}
\def\ox16{\hbox{$^{16}$O}}
\def\oxygen16{\hbox{$^{16}$O}}

\def\phib8{\hbox{$\phi(\b8)$}}
\def\phibe7{\hbox{$\phi(\be7)$}}

\def\phin13{\hbox{$\phi(\n13)$}}
\def\phio15{\hbox{$\phi(\o15)$}}

\def\sec-1{\hbox{sec$^{-1}$}}
\def\sec{\hbox{\rm sec}\tcar}

\def\sn115{\hbox{$^{115}$Sn}}

\def\tc98{\hbox{$^{98}$Tc}}
\def\tl205{\hbox{$^{205}$Tl}}

\def\va51{\hbox{$^{51}$V}}

\def\xenon127{\hbox{$^{127}$Xe}}

\def\zn65{\hbox{$^{65}$Zn}}

\def\ea37{\a37}
\def\ear37{\ar37}
\def\eargon40{\argon40}
\def\eb8{\b8}
\def\ebe7{\be7}
\def\eberil8{\beril8}
\def\eboron11{\boron11}
\def\ebr81{\br81}
\def\ec12{\carb12}
\def\eca37{\ca37}
\def\ecar11{\car11}
\def\ecarbon13{\carbon13}

\def\echr51{\chr51}
\def\ecl37{\cl37}

\def\ecu65{\cu65}

\def\eeminus{\eminus}
\def\eeminus\rm {e^-}

\def\ega71{\ga71}
\def\ege71{\ge71}

\def\eh2{\h2}
\def\ehe3{\he3}

\def\ehelium4{\helium4}

\def\ehy1{\hy1}
\def\ein115{\ind115}
\def\ek40{\k40}
\def\ekay40{\kay40}
\def\ekr81{\kr81}
\def\eli7{\li7}
\def\elithium8{\lithium8}

\def\eminus{\hbox{$\rm e^-$}}
\def\emo98{\mo98}
\def\emo98{\mo98}

\def\en13{\n13}
\def\enitrogen14{\nitrogen14}

\def\eo15{\o15}
\def\eoxygen16{\oxygen16}

\def\ephib8{\phi(\b8)}
\def\ephibe7{\phibe7}

\def\ephin13{\phin13}
\def\ephio15{\phio15}

\def\esn115{\sn115}

\def\etc98{\tc98}

\def\etal{{\it et al.}}
\def\eva51{\va51}

\def\ezn65{\zn65}

\def\gtorder{\mathrel{\raise.3ex\hbox{$>$}\mkern-14mu
             \lower0.6ex\hbox{$\sim$}}}
\def\ltorder{\mathrel{\raise.3ex\hbox{$<$}\mkern-14mu
             \lower0.6ex\hbox{$\sim$}}}

\catcode`\@=11 
\newif\iftwelv@  \twelv@true

\def\Textindent#1{\noindent\llap{#1\enspace}\ignorespaces}
\newdimen\referenceminspace  \referenceminspace=20pc
\def\testnextcar{\ifhmode\ifcat\next.\else\ \fi\fi}
\def\tcar{\futurelet\next\testnextcar}
\let\rel@x=\relax
\let\n@expand=\relax
\def\pr@tect{\let\n@expand=\noexpand}
\let\protect=\pr@tect
\let\gl@bal=\global
\newdimen\d@twidth
{\setbox0=\hbox{s.} \gl@bal\d@twidth=\wd0 \setbox0=\hbox{s}
        \gl@bal\advance\d@twidth by -\wd0 }
\def\removehglue{\loop \unskip \ifdim\lastskip >\z@ \repeat }
\def\roll@ver#1{\removehglue \nobreak \count255 =\spacefactor \dimen@=\z@
        \ifnum\count255 =3001 \dimen@=\d@twidth \fi
        \ifnum\count255 =1251 \dimen@=\d@twidth \fi
    \iftwelv@ \kern-\dimen@ \else \kern-0.83\dimen@ \fi
   #1\spacefactor=\count255 }
\def\step@ver#1{\rel@x \ifmmode #1\else \ifhmode
        \roll@ver{${}#1$}\else {\setbox0=\hbox{${}#1$}}\fi\fi }
\def\attach#1{\step@ver{\strut^{\mkern 2mu #1} }}
%
%
\def\spacecheck#1{\dimen@=\pagegoal\advance\dimen@ by -\pagetotal
   \ifdim\dimen@<#1 \ifdim\dimen@>0pt \vfil\break \fi\fi}
%
%
%
%
\newcount\referencecount     \referencecount=0
\newcount\lastrefsbegincount \lastrefsbegincount=0
\newif\ifreferenceopen       \newwrite\referencewrite
\newdimen\refindent          \refindent=30pt
\def\normalrefmark#1{\attach{\scriptstyle  #1  }}

\def\NPrefmark#1{\step@ver{{\;[#1]}}}
\def\refmark#1{\unskip\normalrefmark{#1}}
\def\refend@{\refmark{\number\referencecount}}
\def\refend{\refend@{}\space }
\def\refsend{\refmark{\count255=\referencecount
   \advance\count255 by-\lastrefsbegincount
   \ifcase\count255 \number\referencecount
   \or \number\lastrefsbegincount,\number\referencecount
   \else \number\lastrefsbegincount-\number\referencecount \fi}\space }
\def\REFNUM#1{\gl@bal\advance\referencecount by 1
    \xdef#1{\the\referencecount }}
\def\Refnum#1{\REFNUM #1\refend@ } 
\def\REF#1{\REFNUM #1\R@FWRITE\ignorespaces}
\def\Ref#1{\Refnum #1\REFWRITE }
\def\ref{\Ref\?}
\def\REFS#1{\REFNUM #1\gl@bal\lastrefsbegincount=\referencecount
    \REFWRITE }

       \let\REFSCON=\REF
\def\rf#1{\Ref#1}
\def\rfs#1{\REFS#1}
\def\rfscon#1{\REFSCON#1}
\def\r@fitem#1{\par \hangafter=0 \hangindent=\refindent \Textindent{#1}}
\def\refitem#1{\r@fitem{#1.}}
\def\NPrefitem#1{\r@fitem{[#1]}}
\def\NPrefs{\let\refmark=\NPrefmark \let\refitem=NPrefitem}
\begingroup
 \catcode`\^^M=\active \let^^M=\relax %
 \gdef\p@rse@ndwrite#1#2{\begingroup \catcode`\^^M=12 \newlinechar=`\^^M%
         \chardef\rw@write=#1\sc@nlines#2}%
 \gdef\sc@nlines#1#2{\sc@n@line \g@rbage #2^^M\endsc@n \endgroup #1}%
 \gdef\sc@n@line#1^^M{\expandafter\toks@\expandafter{\deg@rbage #1}%
         \immediate\write\rw@write{\the\toks@}%
         \futurelet\n@xt \sc@ntest }%
\endgroup
\def\sc@ntest{\ifx\n@xt\endsc@n \let\n@xt=\rel@x
       \else \let\n@xt=\sc@n@notherline \fi \n@xt }
\def\sc@n@notherline{\sc@n@line \g@rbage }
\def\deg@rbage#1{}
\let\g@rbage=\relax    \let\endsc@n=\relax
\def\REFWRITE{\R@FWRITE\rel@x }
\def\R@FWRITE#1{\ifreferenceopen \else \gl@bal\referenceopentrue
     \immediate\openout\referencewrite=\jobname.refs
     \toks@={\begingroup \refoutspecials \catcode`\^^M=10 }%
     \immediate\write\referencewrite{\the\toks@}\fi
    \immediate\write\referencewrite{\noexpand\refitem %
                                    {\the\referencecount}}%
    \p@rse@ndwrite \referencewrite #1}
\def\refout{\vfill\eject
   \spacecheck\referenceminspace
   \ifreferenceopen \Closeout\referencewrite \referenceopenfalse \fi
   \line{\hfil REFERENCES\hfil}\vskip.25in
   \input \jobname.refs
   }
\def\refoutspecials{\sfcode`\.=1000 \interlinepenalty=1000
         \rightskip=\z@ plus 1em minus \z@ }
\def\Closeout#1{\toks0={\par\endgroup}\immediate\write#1{\the\toks0}%
   \immediate\closeout#1}
\catcode`\@=12 
\def\AB{\bigskip\parindent=40pt
        \centerline{\bf ABSTRACT}\medskip\halfspace\narrower}
\def\AE{\bigskip\nonarrower\doublespace}
\def\nonarrower{\advance\leftskip by-\parindent
        \advance\rightskip by-\parindent}

\centerline{\bf DO SOLAR NEUTRINO EXPERIMENTS IMPLY NEW PHYSICS?}
\bigskip
\centerline{John N. Bahcall}
\smallskip
\centerline{Institute for Advanced Study, Princeton, New Jersey}
\medskip
\centerline{H. A. Bethe}
\smallskip
\centerline{Newman Laboratory of Nuclear Studies, Cornell
University, Ithaca, NY}
\smallskip
\centerline{Received August 1992}
\doublespace
\AB
None of the
1000 solar models in a full Monte Carlo simulation is consistent with the
results of the
chlorine or the Kamiokande experiments.
  Even if the
solar models are forced artifically
 to have a \b8 neutrino flux in agreeement with the Kamiokande
experiment, none of the fudged models agrees
 with the chlorine observations.
The GALLEX and SAGE experiments, which
currently have large statistical
uncertainties, differ from the predictions of the standard solar model
by $2 \sigma$ and $3 \sigma$, respectively.

\AE
\bigskip
Four solar neutrino experiments
\rfs\Davis{
R. Davis Jr., in {\it Proc. of Seventh Workshop on Grand
Unification, ICOBAHN'86}, edited by J. Arafune (World
Scientific, Singapore, 1987),
p. 237; R. Davis Jr., K. Lande, C. K. Lee, P. Wildenhain, A.
Weinberger, T.  Daily, B. Cleveland, and J. Ullman, in {\it Proceedings of
the 21st International Cosmic Ray Conference}, Adelaide, Australia,
1990,
in press; J. K. Rowley, B. T. Cleveland, and R.
Davis Jr., in {\it Solar
Neutrinos and Neutrino Astronomy}, edited by  M. L. Cherry, W. A. Fowler, and
K. Lande (American Institute of Physics, New York, 1985), Conf. Proceeding No.
126, p. 1.}
\rfscon\Hirata{K. S. Hirata {\it et al.},
Phys. Rev. Lett. {\bf 63}, 16 (1989); {\bf 65}, 1297 (1990).}
\rfscon\SAGEone{A. I. Abazov {\it et al.},
Phys. Rev. Lett. {\bf 67}, 332 (1991).}
\rfscon\GALLEXone{P. Anselmann {\it et al.}, Phys. Lett. B
{\bf 285}, 376 (1992).}
\refsend
yield results different from the
combined predictions of the standard solar model and the standard
electroweak model with zero neutrino masses.
The question physicists most often ask each other about these results
is:
Do these experiments require new physics beyond the
standard electroweak model?
We provide a quantitative answer using the results of a
detailed Monte Carlo study of the predictions of the standard solar
model.

The basis for our investigation is a collection of 1000
precise solar models \rf\Bahcall{J. N. Bahcall and R. K. Ulrich,
Rev. Mod. Phys. {\bf 60}, 297 (1988);
J. N. Bahcall, {\it Neutrino Astrophysics}
(Cambridge University Press, Cambridge, England, 1989).}
in which each input parameter
(the principal nuclear reaction
rates, the solar composition, the solar age, and the radiative opacity)
for each model was drawn randomly
from a normal distribution with the mean and standard deviation
appropriate to that variable.
In the calculations described in this paper,
the uncertainties in the neutrino cross sections \attach{\Bahcall}
for chlorine and for gallium were included by assuming a normal
distribution for each of the absorption cross sections with its estimated
mean and error. Since for gallium the estimated uncertainties in
the neutrino absorption
cross sections are not symmetric
with respect to the
best-estimate absorption cross sections, two different
normal distributions were
used to simulate the detection uncertainties for each neutrino flux
to which gallium is sensitive.
This Monte Carlo study automatically
takes account of the nonlinear relations
among the different neutrino fluxes that are imposed by the coupled
partial differential equations of stellar structure
and by
the boundary conditions of matching the observed solar luminosity, heavy
element to hydrogen ratio, and effective temperature at the present
solar age.
This is the first Monte Carlo study that uses large numbers of standard
solar models that satisfy the equations of stellar evolution and
that is designed to determine if physics
beyond the standard electroweak theory is required.

Related investigations have been carried out assuming
\rfs\Bludmanone{S. A. Bludman, D. C. Kennedy, and P. G. Langacker,
Nucl. Phys. B {\bf 374}, 373 (1992).}
\rfscon\GALLEXONE{P. Anselmann \etal, Phys. Letter B {\bf 285},
390 (1992).}
\rfscon\BludmanTwo{S. A. Bludman, N. Hata, D. C. Kennedy, and P. G.
Langacker, University of Pennsylvania preprint UPR-0516T (1992).}
\refsend
that each solar
model could be represented by a single parameter,
its central temperature.
The flux of neutrinos from each nuclear source is represented in these
studies by a power
law in the central temperature.
This simplification, while providing a semi-quantitative understanding
of some of the important relationships,
 leads to serious errors in some cases.
For example, the parameterization in terms of central temperature
predicts a \b8 flux for the Maximum Rate Model that is too low by more
than a factor of four \rf\Yalepaper{J. N. Bahcall and M.H. Pinsonneault,
Rev. Mod. Phys. (1992), in press.}.
Just as  detailed Monte Carlo calculations are necessary in order to understand
the relative and absolute sensitivities of complicated laboratory
experiments,
a full Monte Carlo calculation is required to determine the
interrelations and absolute values
of the different solar neutrino fluxes.  The sun is as complicated as a
laboratory accelerator or a laboratory detector, for which we know by
painful experience that detailed simulations are necessary.
For example, the fact that the
${\rm ^8B}$ flux may be crudely described as $\phi({\rm
^8B})~\propto~T^{18}_{\rm central}$ and $\phi({\rm
^7Be})~\propto~T^{8}_{\rm central}$ does not specify whether the two
fluxes increase and decrease together or whether their changes are out
of phase with each other.  The actual variations of the calculated
neutrino fluxes are determined by the coupled partial differential
equations of stellar evolution and the boundary conditions, especially
the constraint that the model luminosity at the present epoch be equal
to the observed solar luminosity.
Fortunately, the simplified method and the full Monte Carlo
calculation yield similar results for the probability that new physics
is required, although the full Monte Carlo calculation yields  a more
accurate numerical statement.

Figures 1a-c show the number of solar models with different predicted
event rates for the chlorine
solar neutrino experiment, the Kamiokande (neutrino-electron scattering)
experiment, and the two gallium experiments (GALLEX and SAGE).
For the chlorine experiment, which is sensitive to neutrinos above
0.8~Mev,
the solar model with the best input
parameters predicts\attach{\Bahcall} an event rate of about 8 SNU.
None of the 1000 calculated solar models yields a capture
rate below 5.8 SNU, while the observed rate is \attach{\Davis}
\eqnam{\chlorinerate}
$$
<\phi \sigma >_{\rm Cl~exp}  ~=~ (2.2 \pm 0.2) ~ {\rm SNU},~~~
1 \sigma ~{\rm error} .
\eqno\new)
$$
The discrepancy that is apparent in Figure~1a was for two decades the
entire ``solar neutrino problem.''
Figure~1a implies that something is wrong
with either the standard solar model or the standard electroweak
description of the neutrino.

Figure~1b shows the number of solar models with different \b8 neutrino
fluxes.  For convenience, we have divided each \b8 flux by the average
\b8 flux so that the distribution is peaked near 1.0.
The rate measured for neutrinos with energies above 7.5~MeV
by the Kamiokande~II and III experiments is
\attach{\Hirata}

\eqnam{\KIIrate}
$$
<\phi(^8B)  > ~= ~
[  0.48 \pm 0.05 (1 \sigma) \pm 0.06 ({\rm syst}) ]
<\phi(^8B) >_{\rm Average}
\eqno\new)
$$
for recoil electrons with energies greater than 7.5 MeV.
Here $<\phi(^8B) >_{\rm Average}$ is the best-estimate theoretical
prediction. None of the 1000 standard solar models lie
below \hbox{$0.65~<\phi(^8B) >_{\rm Average}$}.
If one takes account of the Kamiokande measurement uncertainty($\pm
0.08$) in the Monte Carlo simulation, one still finds that none of the
solar models are consistent with the observed event rate.
This results provides independent support for the existence of a
solar neutrino problem.

Figure~1c shows the number of solar models with different
predicted event rates for gallium detectors and
the recent measurements by the SAGE
\attach{\SAGEone,}\rf\Gavrin{V. N. Gavrin, \etal, {\it XXVI International
Conference on High Energy Physics}, Dallas, Texas (1992).}
 $(58^{+17}_{-24}~\pm~14 ({\rm syst})~{\rm SNU})$ and GALLEX
$(83~\pm~19 (1 \sigma)~\pm~8 ({\rm syst})~{\rm SNU}$)
collaborations \attach{\GALLEXone}.
With the current large statistical errors,
the results differ from the best-estimate theoretical
value \attach{\Bahcall}
of 132~SNU
by approximately $2~\sigma$ (GALLEX) and $3.5~\sigma$
(SAGE). The gallium results provide modest support for the
existence of a solar neutrino problem, but by themselves do
not constitute a strong conflict with standard theory.

Can the discrepancies between observation and calculation that are
summarized in Figure~1 be resolved by changing some aspect of the solar
model?  We have argued previously
\rf\Betheone{J. N. Bahcall and H. A. Bethe, Phys. Rev. Lett.
{\bf 65}, 2233 (1990).}
that this is difficult to do
because the energy spectrum of any specific neutrino source is unchanged
by the solar environment \rf\spectrum{J. N. Bahcall,
Phys. Rev. D. {\bf 44}, 1644 (1991).}
and because the uncertainties in all of the important sources except
\b8 are relatively small.  A comparison of Figures~1a-1b shows that the
discrepancy with theory appears to be energy dependent.  The larger
discrepancy occurs for the chlorine experiment, which is sensitive to
lower neutrino energies than is the Kamiokande experiment.
If one normalizes the \b8 flux by the best-estimate measurement from
Kamiokande (see \eqt{\KIIrate}), then the implied rate
in the chlorine experiment from \b8 alone is
2.9 SNU .  We also argued that the other neutrino fluxes would most likely
yield at least another 1 SNU, implying a minimum total rate of
about 4 SNU.  On this basis, we concluded that the two experiments--
chlorine and Kamiokande-- are inconsistent with the combined standard
electroweak and solar models.
However, our argument did not take into account in a well-defined way
the errors in the
predictions and in the measurements.  We remedy this shortcoming in the
following discussion using the previously-described Monte Carlo
simulation.

Figures~2 provides a quantitative expression of the
difficulty in reconciling the Kamiokande and chlorine experiments by
changing solar physics.
We constructed Figure~2 using the same 1000 solar models as were used in
constructing Figure~1, but for Figure~2 we artificially replaced the \b8 flux
for each standard model by a value drawn randomly
for that model from a normal distribution with
the mean and the standard deviation measured by Kamiokande (see
\eqt{\KIIrate}).  This {\it ad hoc} replacement is motivated by the fact
that the $^7$Be~$(p, \gamma)$~$^8$B cross section is the least accurately
measured of all the relevant nuclear fusion cross sections and by the
remark that the $^8$B
neutrino flux is more sensitive to solar interior conditions than any of
the other neutrino fluxes.  The peak of the resulting distribution is moved
 to 4.7 SNU (from 8 SNU) and the full width of the peak is
decreased by about a factor of three.
The peak is displaced because the measured (i.e., Kamiokande) value
of the \b8 flux is smaller than the calculated value.
 The width of the distribution is decreased because
the error in the Kamiokande measurement is less than the
estimated theoretical uncertainty ($\approx 12.5\%$)
and because \b8 neutrinos constitute a smaller fraction of each displaced
rate than of the corresponding standard rate.

Figure~2 was constructed by assuming that something is seriously wrong
with the standard solar model, something that is
sufficient to cause the \b8 flux to be
reduced to the value measured in the Kamiokande experiment.
Nevertheless,
there is no overlap between the distribution of fudged standard
model rates and the measured chlorine rate.  None of
the 1000 fudged models lie within $3 \sigma$ (chlorine measurement
errors) of the
experimental result.

The results presented in Figures~1-2
suggest that new physics is required beyond the standard
electroweak theory if
the existing solar neutrino
experiments are correct within their quoted uncertainties.
Even if one abuses the solar models by artifically imposing
consistency with the
Kamiokande experiment, the resulting predictions of all 1000 of
the ``fudged'' solar models are inconsistent with the result of the
chlorine experiment(see Figure~2).  Figure~3 shows the relatively high
precision with which the $pp$ and ${\rm ^7Be}$ neutrino fluxes can be
calculated.  Since the $1 \sigma$ theoretical
uncertainty in the flux of $^7$Be neutrinos is only $\simeq~5\%$
\attach{\Bahcall,\Yalepaper}, an accurate measurement of this quantity
in the proposed Borexino experiment \rf\Raghavan{R. S. Raghavan, in
{\it Proc.
Int. Conf. High Energy Physics},
Singapore, edited by  K. K. Phua and Y. Yamaguchi (World Scientific,
Singapore, 1990), Vol. 1, p. 482; C. Arpasella \etal, in {\it
Borexino at Gran Sasso:
Proposal for a real-time detector for low energy solar neutrinos},
Vols. I and II, University of Milan, INFN Report.}
 will constitute another important test of the standard model.

All of the arguments in this paper depend to some extent on our
understanding of the solar interior.
In the future, it will be possible
to use solar neutrinos to test electroweak theory
independent of solar models by measuring the energy spectrum of the
\b8 neutrinos with the Super-Kamiokande \rf\SuperK{Y. Totsuka,
in {\it Proc.
of the International Symposium on Underground Physics Experiments},
edited by  K. Nakamura (ICRR, Univ. of Tokyo, 1990), p. 129.},
the SNO\rf\SNO{G. Aardsma \etal, Phys. Lett. B {\bf 194}, 321 (1987).}, and the
ICARUS experiments \rf\ICARUS{J. N. Bahcall, M. Baldo-Ceollin, D. Cline,
and C. Rubbia, Phys. Lett. B {\bf 178}, 324 (1986).}
 and by measuring the ratio of charged to neutral currents
with the SNO experiment\attach\SNO.
\bigskip

\centerline{ACKNOWLEDGMENTS}

This work was supported in part by the NSF via grant
PHY-91-06210 at I.A. S. and PHY-87-15272 at Cornell.

\Rf

\refout
\vfill
\eject
\bigskip
\doublespace
\centerline{FIGURE CAPTIONS}

\noindent
FIG. 1.\ 1000 Solar Models versus Experiments.  The number of
precisely-calculated
solar models that predict different solar neutrino event rates are shown
for the chlorine (Figure~1a), Kamiokande (Figure~1b), and gallium
(Figure~1c) experiments.  The solar models from which the fluxes were
derived satisfy the equations of stellar evolution including the
boundary conditions that the model luminosity, chemical composition, and
effective temperature at the current solar age be equal to the observed
values \attach{\Bahcall}.  Each input parameter in each solar model was drawn
independently from a normal distribution having the mean and the
standard deviation appropriate to that parameter.  The experimental
error bars include both statistical errors ($1 \sigma$) and systematic
uncertainties, combined linearly.

\noindent
FIG. 2.\ 1000 Artifically Modified Fluxes.  The $^8$B neutrino
fluxes computed for the 1000 accurate solar models were replaced in the
figure shown by values drawn randomly for each model from a normal
distribution with the mean and the standard deviation measured by the
Kamiokande experiment \attach{\Hirata}.

\noindent
FIG. 3.\ The $pp$ and ${\rm ^7Be}$ Neutrino Fluxes.  The histogram of
the number of 1000 precisely-calculated solar models that predict different
$pp$ neutrino fluxes is shown in Figure~3a and the number that predict
different ${\rm ^7Be}$ neutrino fluxes is shown in Figure~3b.  The
individual neutrino fluxes are divided by their respective average
values.
\bye